\documentclass[preprint2,times,tighten]{aastex6}
\pdfoutput=1 %for arXiv submission
\usepackage{amsmath,amstext}
\usepackage[T1]{fontenc}
\usepackage[figure,figure*]{hypcap}

\shorttitle{Hot Start Formation Models}
\shortauthors{Berardo et al.}

\begin{document}

\title{Hot Start Giant Planets Form With Radiative Interiors}

\author{David Berardo\altaffilmark{1,2}, Andrew Cumming\altaffilmark{1,2}}
\affil{
\altaffilmark{1}{Department of Physics and McGill Space Institute, McGill University, 3600 rue University, Montreal, QC, H3A 2T8, Canada}\\
\altaffilmark{2}{Institut de recherche sur les exoplan\`{e}tes (iREx)}
}
\email{david.berardo@mcgill.ca}
\email{andrew.cumming@mcgill.ca}

\begin{abstract}
In the hot-start core accretion formation model for gas giants, the interior of a planet is usually assumed to be fully convective. By calculating the detailed internal evolution of a planet assuming hot start outer boundary conditions, we show that such a planet will in fact form with a radially increasing internal entropy profile, so that its interior will be radiative instead of convective. For a hot outer boundary, there is a minimum value for the entropy of the internal adiabat $S_{min}$ below which the accreting envelope does not match smoothly onto the interior, but instead deposits high entropy material onto the growing interior. One implication of this would be to at least temporarily halt the mixing of heavy elements within the planet, which are deposited by planetesimals accreted during formation. The compositional gradient this would impose could subsequently disrupt convection during post-accretion cooling, which would alter the observed cooling curve of the planet. However even with a homogeneous composition, for which convection develops as the planet cools, the difference in cooling timescale will change the inferred mass of directly-imaged gas giants.
\end{abstract}


\section{Introduction}

Giant planets may form from core accretion, in which runaway gas accretion occurs onto a $\sim 10\ M_\earth$ core, or from direct collapse from the gas disk (see \citealt{Helled2014} for a review). A number of observational constraints on how gas giant planets form are becoming available, both in our Solar System and in exoplanetary systems. Young massive giant planets have been directly imaged \citep{Bowler2016}, revealing their thermal state $\sim 10^7\ {\rm yrs}$ after formation and the composition of their atmospheres. Exoplanet surveys have measured occurrence rates and orbital architectures of planetary systems containing gas giants (e.g.~\citealt{Clanton2016}). In the Solar System, a recent example is the precise measurements of Jupiter's gravitational moments by {\em Juno} \citep{Bolton2017}, suggesting the core may be dilute, expanded to $\gtrsim 0.3$ of Jupiter's radius \citep{Wahl2017}. This indicates either that the core can be mixed upwards during evolution, or is telling us about the distribution of heavy elements at formation. This variety of observations motivate continued theoretical work on the physics of gas giant formation. 

A major uncertainty in the core accretion scenario is the efficiency of the shock that forms at the surface of the planet during runaway accretion \citep{Marley2007}. Depending on how much of the gravitational energy of the infalling matter is radiated away at the shock, the luminosity of the planet after formation can differ by orders of magnitude, leading to uncertainty in derived planet masses \citep{Marley2007,Spiegel2012,Marleau2014}. Recent work has suggested, however, that a hot start is more likely than a cold start. \cite{Marleau2017} carried out 1D radiation-hydro simulations of the shock and found that a significant fraction of the gravitational energy is incorporated into the planet (see also \citealt{Szulagyi2017}). \cite{Berardo2017} studied the growth of giant planets treating the shock temperature as a free parameter. They found that the cold starts of \cite{Marley2007} (based on the simulations of \citealt{Hubickyj2005}) required very low boundary temperatures: close to the disk temperature, and lower than the photospheric temperature of the planet. \cite{Owen2016} studied growth by disk accretion and found hot starts when the boundary layer thickness exceeded a critical  value.

In this paper, we present detailed models of the runaway accretion phase of gas giant growth under the assumption of a hot start. 
Previous core accretion models by \cite{Pollack1996}, \cite{Bodenheimer2000}, \cite{Hubickyj2005}, and \cite{Lissauer2009} assumed cold outer boundaries. \cite{Mordasini2013} calculated hot start models by stepping through pre-computed planet models that assumed a constant internal luminosity \citep{Mordasini2012} and so did not follow the effect of accretion on the internal structure. Here, we use the Modules for Experiments in Stellar Astrophysics (MESA) code \citep{Paxton2011,Paxton2013,Paxton2015} to calculate the internal structure during accretion with a hot start boundary condition. We show that the planet forms in successive layers of increasing entropy\footnote{In this paper, we use the term entropy to refer to the specific entropy, measured in units of $k_B/m_p$, where $k_B$ is Boltzmann's constant and $m_p$ is the proton mass.}, inhibiting convection and giving a radiative interior (this possibility was discussed by \citealt{Mordasini2012} based on previous work on accretion onto low mass stars, e.g.~\citealt{Prialnik1985}). In \S 2, we discuss the entropy of matter deposited by the accreting envelope and show that the evolution of the shock temperature with time determines whether the growing planet is convective or radiative. In \S 3, we present numerical models with MESA that follow the planet growth and subsequent cooling. We discuss the implications of our results in \S4.

\section{The Entropy of Matter Deposited by the Accreting Envelope}

During runaway gas accretion, infalling matter is decelerated at an accretion shock at the planet's surface \citep{Bodenheimer2000}. The post-shock conditions depend on how much of the accretion energy is radiated away at the shock; this sets the post-shock pressure $P_0$ and temperature $T_0$ (e.g.~\citealt{Marleau2017}). \cite{Berardo2017} studied the subsequent evolution of the accreting matter as it settles into the envelope of the planet. They showed that for sufficiently large $T_0$, the radiative envelope is not able to accommodate the large contrast in entropy between the post-shock matter with entropy $S_0$ and the interior with entropy $S_i$ (a similar result was found for accreting protostars by \citealt{Stahler1988}). The entropy in the envelope decreases to a minimum value $S_{\rm min}>S_i$, and the accreting envelope effectively deposits matter with entropy $S_{\rm min}$ onto the growing interior. This contrasts with lower values of $T_0$ for which the entropy in the radiative envelope decreases from $S_0$ to $S_i$, and joins smoothly onto the interior profile (\citealt{Berardo2017} referred to this as the ``stalling'' regime as the cooling rate of the interior is slowed under these conditions).

\begin{figure}
\epsscale{1.2}
\plotone{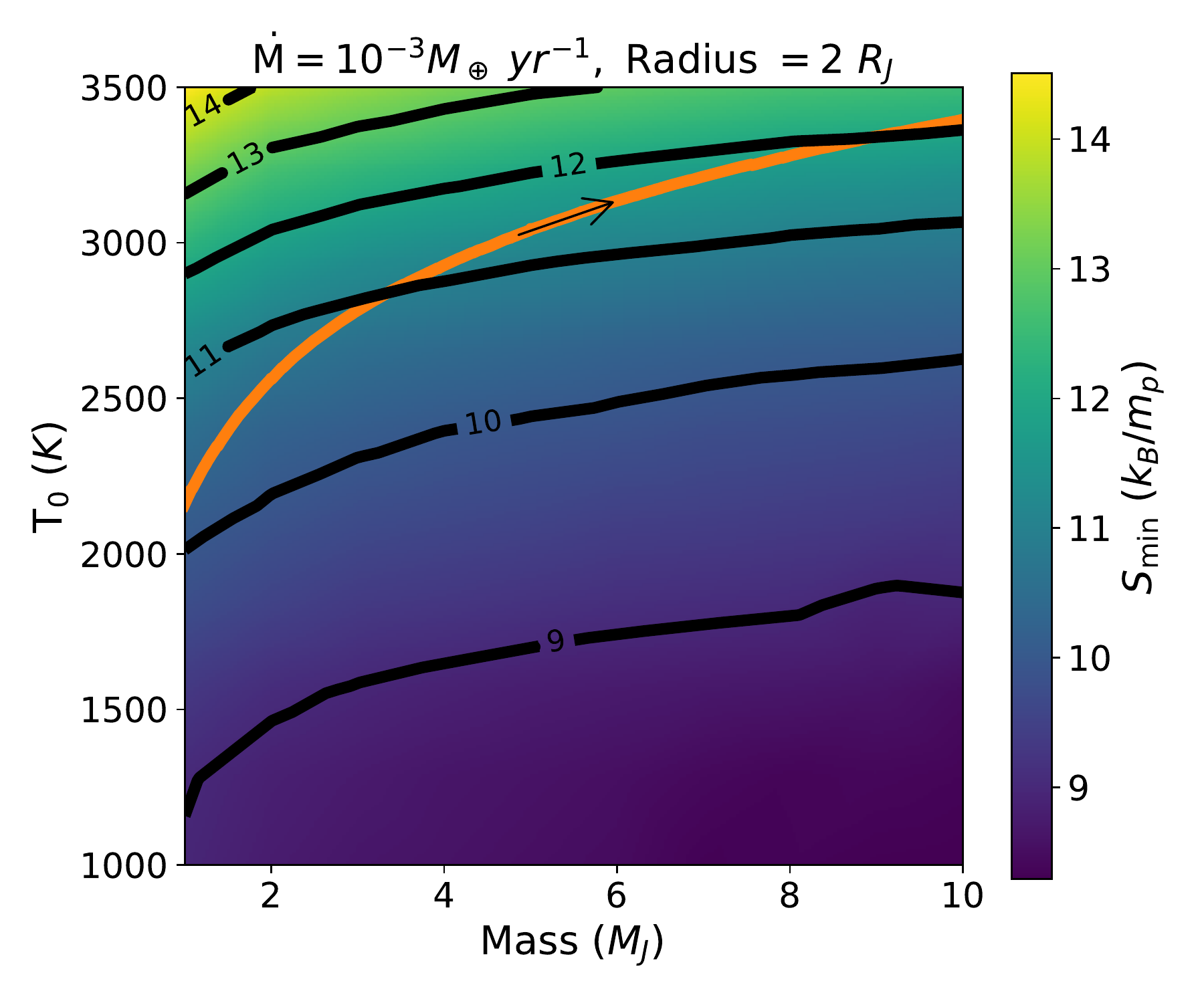}
\caption{The entropy deposited at the base of the accreting envelope $S_{\rm min}$ as a function of shock temperature $T_0$ and planet mass $M$. We assume $\dot{M} = 10^{-3}\ M_\oplus\ {\rm yr}^{-1}$ and radius $R=2\ R_J$. The black lines show contours of $S_{\rm min}$. The orange curve shows the trajectory of $T_0$ and $M$ from a time-dependent model (\S 3), with the direction indicated by a black arrow. Since $S_{\rm min}$ increases along the orange curve, the growing planet has a radiative interior.}
\label{fig: mins contour plot}
\end{figure}

The entropy $S_{\rm min}$ depends on the boundary temperature $T_0$, the accretion rate $\dot M$, and the planet mass $M$ and radius $R$. Figure \ref{fig: mins contour plot} shows the value of $S_{\rm min}$ as a function of $T_0$ and $M$ for typical values of $\dot M$ and $R$. We calculate $S_{\rm min}$ as described in \cite{Berardo2017} \footnote{Code available at 
\url{https://github.com/andrewcumming/gasgiant}.}. We construct steady-state models of the accreting envelope, successively lowering the luminosity at the surface until the luminosity at the base of the envelope goes to zero. The entropy at the base of this lowest luminosity envelope is $S_{\rm min}$.

Figure \ref{fig: mins contour plot} shows that, depending on how $T_0$ changes as the planet increases in mass, $S_{\rm min}$ could either increase or decrease over time, which has implications for the internal structure. If $S_{\rm min}$ decreases with time, low entropy matter is deposited on top of high entropy matter. This situation is unstable to convection, and so we expect the growing planet to have a convective interior. If $S_{\rm min}$ increases with time, the planet grows in layers of successively increasing entropy, inhibiting convection and resulting in a radiative interior.

How do we expect $T_0$ to evolve as the planet grows? \cite{Berardo2017} assumed constant $T_0$ during accretion to assess how the choice of $T_0$ affected the outcome of accretion. Figure \ref{fig: mins contour plot} shows that in that case $S_{\rm min}$ decreases with increasing planet mass, so that the growing planet is always convective. However, in reality we expect the shock temperature to evolve as accretion proceeds. The surface temperature can be written \citep{Mordasini2013}
\begin{equation}\label{eq:T0}
	T_0^4 = \frac{1}{4 \pi R^2 \sigma}\left(L_p\  + \eta L_{acc} \right),
\end{equation}
where $L_p$ is the internal luminosity and $L_{\rm accr}$ the accretion luminosity $L_{\rm accr}=GM\dot M/R$. The parameter $\eta$ measures the efficiency with which the shock radiates \citep{Prialnik1985,Hartmann1997,Mordasini2013}. If the shock radiates away all of the accretion luminosity then $\eta=0$, corresponding to a cold start. If instead the accretion energy is not radiated away but advected into the planet, we have $\eta=1$ and a hot start. Assuming $L_{\rm accr}\gg L_p$, the hot start temperature is
\begin{equation}
\label{eq: hot temperature}
	T_0 \approx 1850\ {\rm K}\ \left({\dot M\over 10^{-3}\ M_\oplus\ {\rm yr}^{-1}}\right)^{1/4}\left({M\over M_J}\right)^{1/4} \left({R\over 2R_J}\right)^{-3/4},
\end{equation}
where we scale to the values of $\dot M$ and $R$ in Figure \ref{fig: mins contour plot}. We see from Figure \ref{fig: mins contour plot} that the corresponding value of $S_{\rm min}$ is $\approx 9.5\ k_B/m_p$. Provided that the internal entropy of the planet at the onset of runaway accretion is $S_i\lesssim 9.5\ k_B/m_p$, the accretion will be in the hot regime.

As the planet grows in mass, and assuming $L_{\rm accr}\gg L_p$, equation (\ref{eq: hot temperature}) gives $T_0\propto M^{1/4}R^{-3/4}$, or
\begin{equation}
\label{eq: temp change}
\frac{d\ln T_0}{d\ln M} = \frac{1}{4} - \frac{3}{4}\frac{d\ln R}{d\ln M}.
\end{equation}
A curve of constant $S_{\rm min}$ on the other hand has $\left.d\ln T_0/d\ln M\right|_{S_{\rm min}}\approx 0.1$ (the slope of the black contours in Fig.~\ref{fig: mins contour plot}). We see that as long as $d\ln R/d\ln M$ is not too large ($\lesssim 0.2$), so that $d\ln T_0/d\ln M>\left.d\ln T_0/d\ln M\right|_{S_{\rm min}}$, $S_{\rm min}$ will increase over time. We show in the next section that this is indeed the case in time-dependent models, so that the interior of the forming giant planet is radiative.

\section{Time-dependent simulations of hot starts}

We use the MESA stellar evolution code (\citealt{Paxton2011,Paxton2013,Paxton2015}; version 8118) to compute a time-dependent model of an accreting gas giant with hot-start boundary conditions. We start with a 0.2 $M_J$ planet with internal entropy $S_i=9.5\ k_B/m_p$ (guided by the models of \citealt{Mordasini2013}), hydrogen, helium, and metal fractions of 0.73, 0.25, 0.02 respectively and a 10 $M_\earth$ core of density $10\ {\rm g\ cm^{-3}}$. We accrete at a constant rate of $10^{-3}\ M_\earth\ {\rm yr^{-1}}$ \citep{Lissauer2009} until the planet reaches $10\ M_J$. During accretion, we set the surface pressure to the sum of the ram pressure and photospheric pressure,
\begin{equation}
P_0 =  \frac{\dot{M}}{4\pi R^2}\left(\frac{2GM}{R} \right)^{1/2} + P_{\rm photo}
\end{equation}
\citep{Mordasini2012}, and the temperature $T_0$ according to equation (\ref{eq:T0}) with $\eta=1$. To avoid convergence issues associated with the onset of accretion, we ramp up the surface temperature linearly from its initial value in the $0.2\ M_J$ model to $T_0$ during accretion of the first $0.2\ M_J$ (the first $\approx 6\times 10^4\ {\rm yr}$).

\subsection{Evolution of the shock temperature and radius}

The time evolution of the shock temperature $T_0$ is shown as the orange curve in Figure \ref{fig: mins contour plot}. The radius evolved first as a fully convective object, decreasing as mass increased from an initial value of $2.1\ R_J$. Around a mass of $M=1.5\ M_J$ it reached a minimum of $1.7\ R_J$, and then began to increase as the structure of the planet became predominantly radiative, back to $2\ R_J$ at $M=10\ M_J$. Fitting a power law to the increasing radius gives $d\ln R/d\ln M\approx 0.1$. Equation (\ref{eq: temp change}) then predicts $d\ln T_0/d\ln M\approx 0.15$, which is in good agreement with the increase of $T_0$ with $M$ (a power law fit to the curve in Fig.~\ref{fig: mins contour plot} gives $d\ln T_0/d\ln M\approx 0.18$). 

As discussed in \S 2, when $T_0$ increases with mass steeper than $d\ln T_0/d\ln M\approx 0.1$, we expect $S_{\rm min}$ to increase with time, and the internal structure to be radiative. The increasing radius with mass indicates this. Studies of the response of stars to accretion have shown that whereas fully-convective objects shrink with increasing mass, radiative stars increase in radius, consistent with our results \citep{Prialnik1985,Hjellming1987,Soberman1997}.

\begin{figure}
\epsscale{1.2}
\plotone{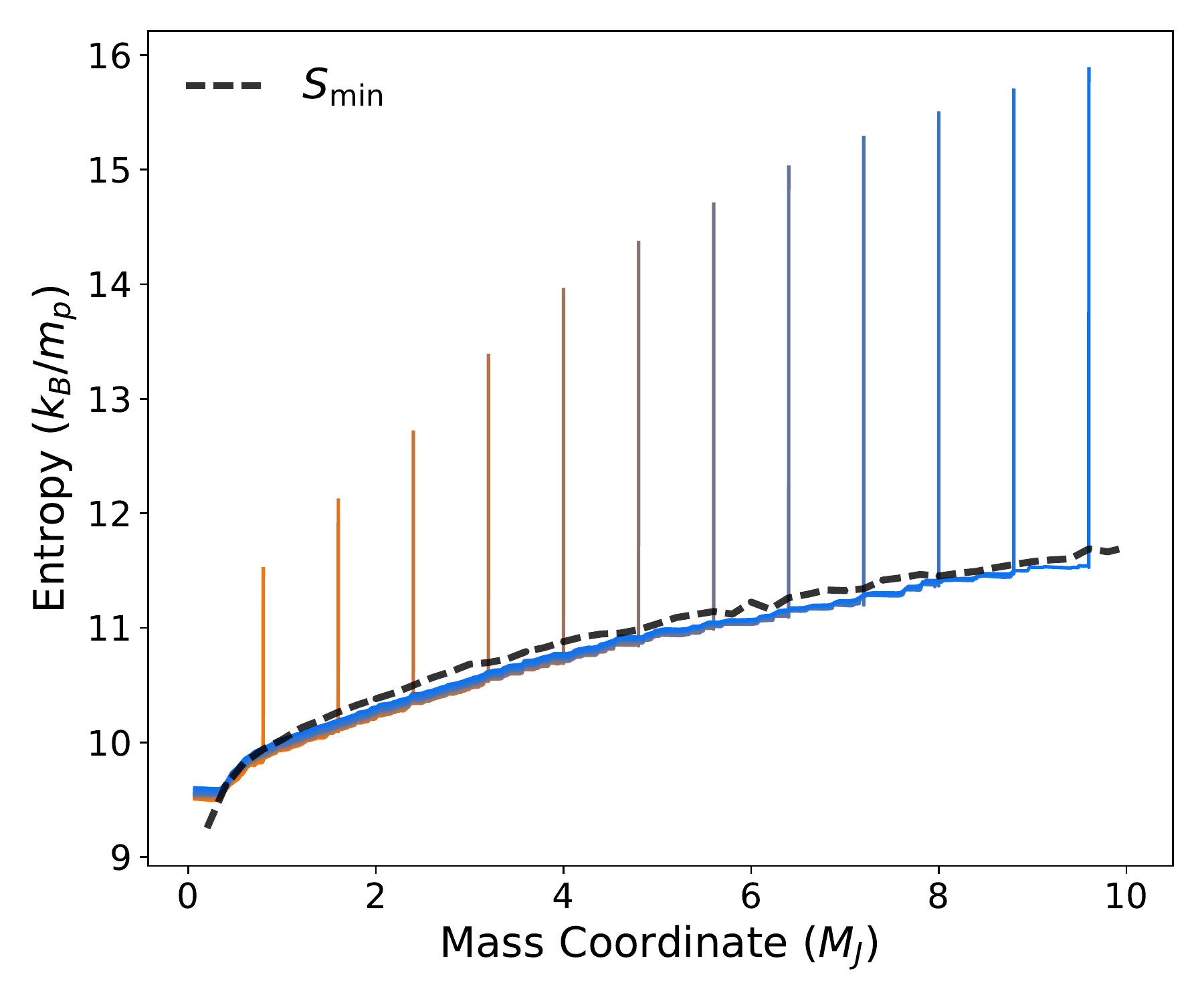}
\caption{Entropy profiles as a function of interior mass coordinate over time. Color indicates the age of the planet, going from orange (youngest) to blue (oldest) as the planet grows to $10\ M_J$. At the assumed constant accretion rate of $\dot M=10^{-3}\ M_\earth\ {\rm yr^{-1}}$, the time to accrete a given mass is $t\approx 3\times 10^5\ {\rm yr}\ (M/M_J)$. The spike in entropy in each model is due to the outer radiative zone at the surface of the planet. The dashed black curve shows the calculated value of $S_\mathrm{min}$ as a function of total mass (i.e. one value of $S_\mathrm{min}$ for every timestep), showing that the internal profile of the planet at the end of accretion is set by the history of $S_\mathrm{min}$ during accretion. }
\label{fig: entropy_vs_mass}
\end{figure}

\begin{figure}
\epsscale{1.2}
\plotone{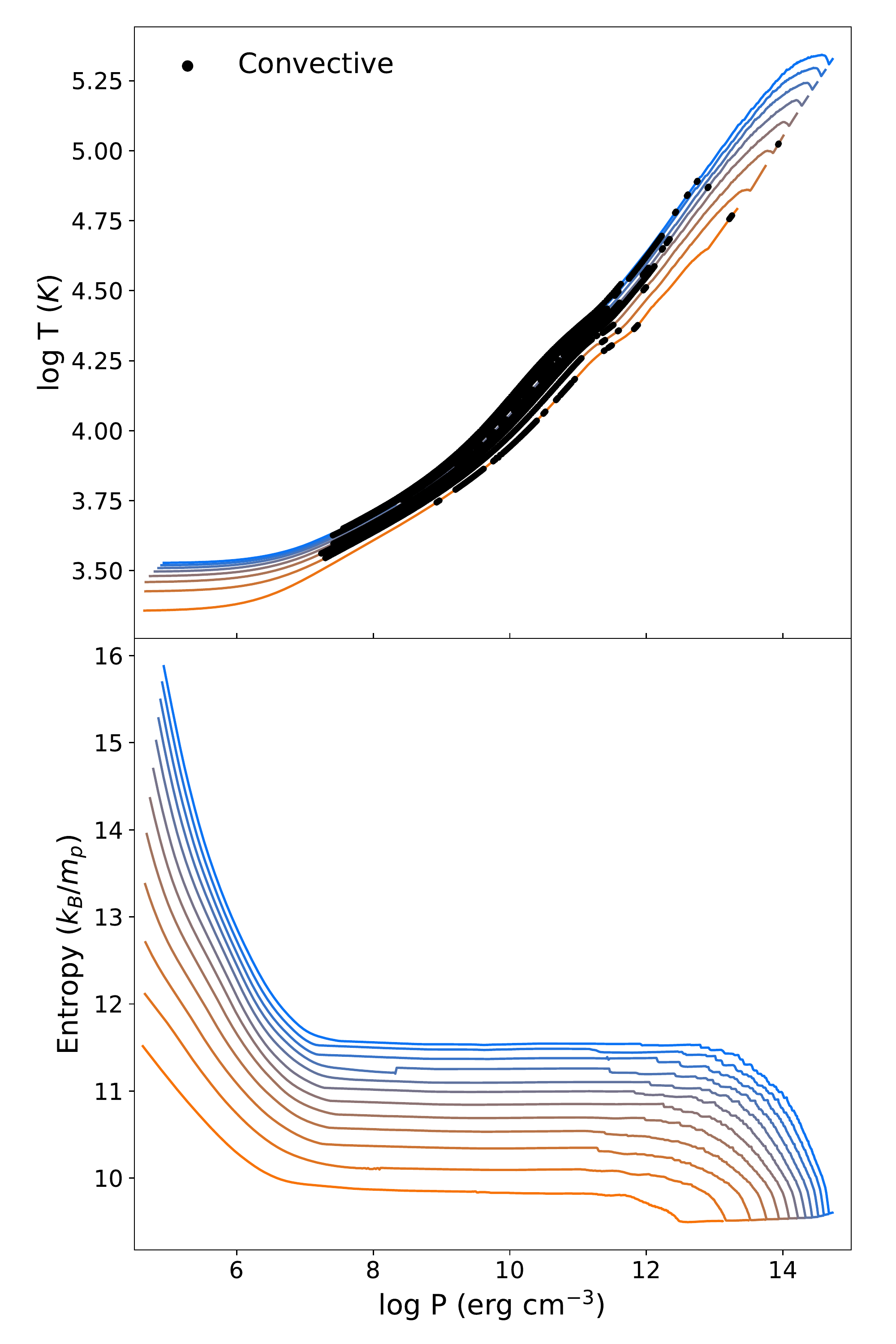}
\caption{Temperature and entropy profiles as a function of pressure during accretion. Color indicates the planet's age as in Fig.~3, going from orange (youngest) to blue (oldest) as the planet grows to $10\ M_J$. In the top panel, black points indicate where the planet is convective according to the Schwarzschild criterion.}
\label{fig: entropy/temp vs pressure}
\end{figure}

\subsection{Internal entropy profile}

Figure \ref{fig: entropy_vs_mass} shows how the entropy profile evolves with time as the mass of the planet grows. The entropy at a given mass coordinate $m$ remains constant as the planet increases in total mass $M$, and the entropy profile is such that entropy increases with $m$, i.e.~increases outwards in the interior.

The increasing entropy profile $S(m)$ is consistent with the expectation from \S 2 that the entropy deposited at the base of the accreting envelope increases over time. To test this idea, we calculated $S_{\rm min}$ as described in \S2 as a function of time (or equivalently total planet mass), using the values of $T_0$, $M$, and $R$ at each timestep. The black dashed curve in Figure \ref{fig: entropy_vs_mass} shows $S_{\rm min}$ as a function of planet mass. We see that it closely reproduces the internal entropy profile, showing that we can understand the growth of the planet as successive layers with entropy $S_{\rm min}$. The timescale for radiative diffusion or thermal conduction is much longer than the accretion timescale, so that the entropy at a given mass coordinate remains unchanged as the planet grows.

Figure \ref{fig: entropy/temp vs pressure} shows the outer envelope in more detail. In the envelope the entropy profile flattens, suggesting the onset of convection. Indeed, we see convection occurring in the envelope, indicated by the solid circles in the top panel of Figure \ref{fig: entropy/temp vs pressure}. However, we note that the convection is irregular, with individual zones switching between convective and radiative as time proceeds (we have checked that this does not depend on our spatial resolution or timestep). The value of entropy at which the envelope flattens corresponds to $S_{\rm min}$. At higher pressures (that make up $\gtrsim$ 99\% of the mass), the structure is radiative, with entropy decreasing to higher pressures.

\begin{figure}
\epsscale{1.2}
\plotone{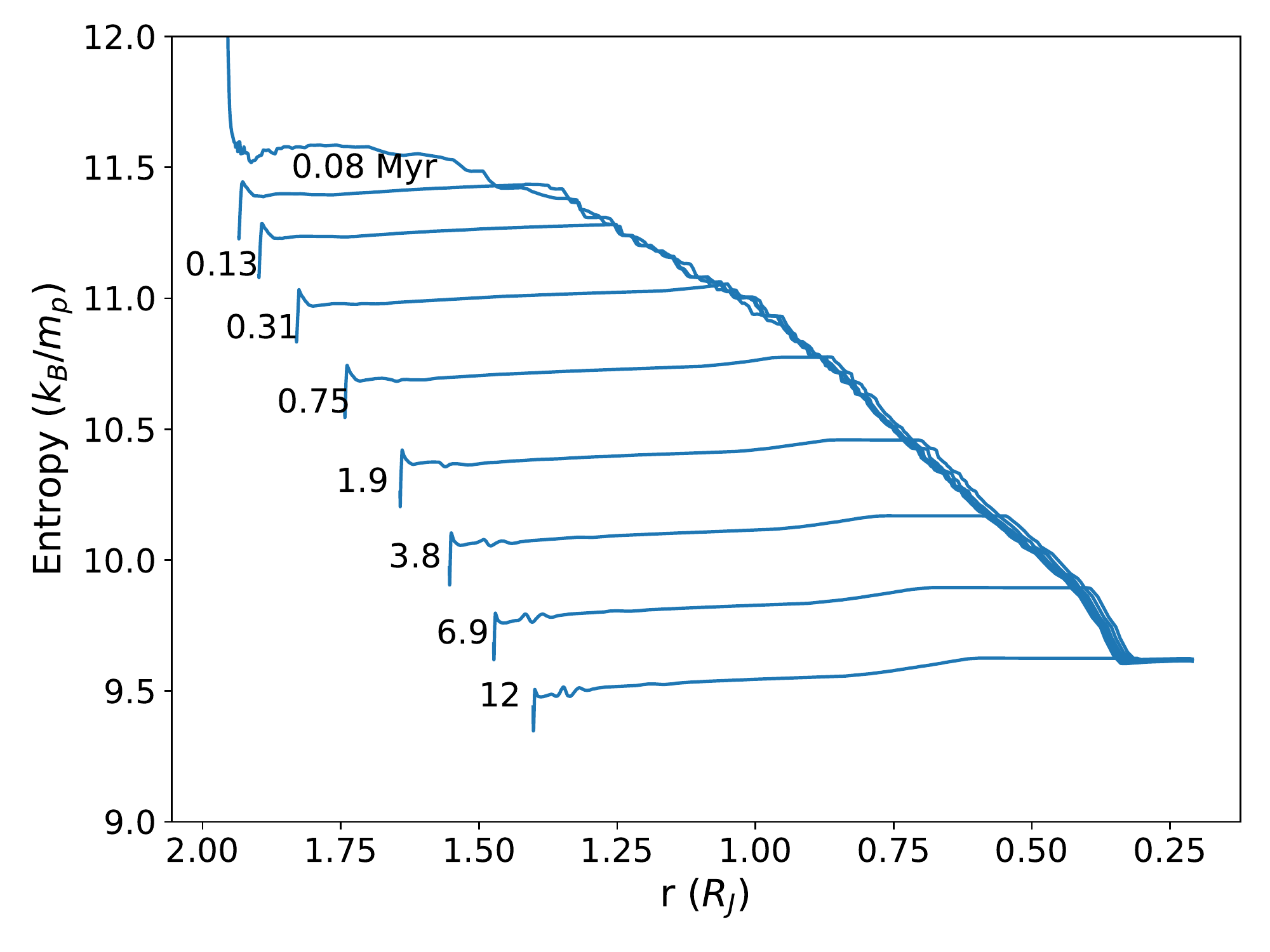}
\caption{The evolution of the internal entropy profile as a function of radial coordinate $r$ as the planet cools post-accretion. The timestamp indicates the elapsed time since the end of accretion. Over time, a convection zone penetrates into the planet, which eventually becomes fully-convective. Towards the surface of each model, the entropy first dips as convection becomes inefficient and then rises again in the surface radiative zone.}
\label{fig: cooling profiles}
\end{figure}

\subsection{Post-Accretion Cooling}

Although the internal structure is radiative during accretion, convection develops once accretion turns off and the planet begins to cool. Figure \ref{fig: cooling profiles} shows the entropy profile at different times following the end of accretion. A convection zone develops at the surface (indicated by the region of constant entropy extending from the surface inwards), and penetrates deeper over time until the whole planet becomes convective. For the $10\ M_J$ planet shown in Figure \ref{fig: cooling profiles}, it takes approximately $10^7\ {\rm years}$ for the planet to become fully convective. For a $1\ M_J$ planet, the timescale is shorter, $\approx 10^6\ {\rm years}$.

\begin{figure}
\epsscale{1.2}
\plotone{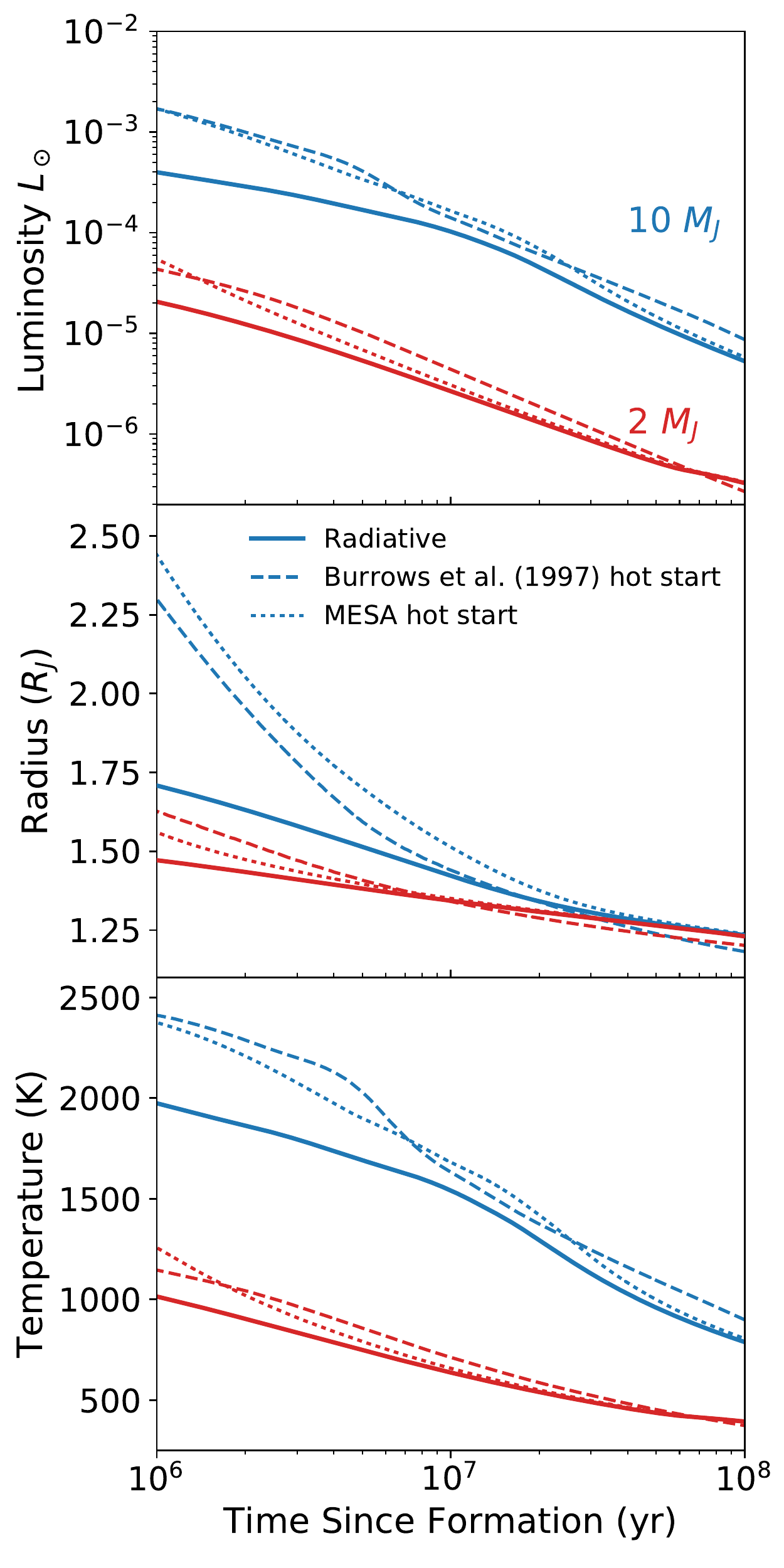}
\caption{Post-accretion cooling models for planets with masses of $2\ M_J$ (red curves) and $10\ M_J$ (blue curves). The solid curves are the radiative models discussed and calculated in this paper. The dotted and dashed curves show hot start cooling models calculated using MESA and from \cite{Burrows1997} respectively.}
\label{fig: cooling curves}
\end{figure}

The timescale for the convection zone to move inwards can be understood in a similar way to cooling of fully-convective planets, by treating the convection zone as a single zone with entropy $S$ (the `following the adiabats' approach, e.g.~\citealt{Fortney2004}). The luminosity leaving the convection zone depends on the opacity at the radiative-convective boundary (RCB) near the surface, and is a function of the entropy and total planet mass, $L(S,M)$. The mass in the convection zone then evolves according to
\begin{equation}\label{eq:dmdt}
	{dM_{\rm conv}\over dt} = {dS/dt\over dS_0/dm} = -{L(S,M)\over M_{\rm conv}\bar{T}dS_0/dm},
\end{equation}
where $dS_0/dm$ is the gradient of the entropy profile $S_0(m)$ at the end of accretion, and $\bar{T}$ is the mass-averaged temperature in the convection zone. The convection zone entropy drops faster than a fully-convective planet because it has a smaller mass and because it is cooler (it occupies the outer regions). 
%As the convection zone penetrates deeper, the cooling timescale becomes closer to the cooling timescale for a planet with the same entropy. 
We have integrated equation (\ref{eq:dmdt}) over time using $L(S,M)$ from \cite{Marleau2014}, and find good agreement with MESA.

Luminosity, radius and effective temperature are shown in Figure \ref{fig: cooling curves}  for two planet masses, and compared to fully-convective hot starts calculated using MESA and from \cite{Burrows1997}. The inwardly decreasing entropy means that the planet is more compact for its luminosity than a fully-convective object. The differences are more pronounced at larger planet masses and earlier times. At $1\ {\rm Myr}$, the luminosity is a factor of $\approx 2$ ($\approx 4$) times smaller and the radius $\approx 6$\% ($\approx 25$\%) smaller than the convective hot start for $M=2\ M_J$ ($10\ M_J$). These differences diminish over time until the planet becomes fully convective at $\sim 10^7\ {\rm years}$.

\section{Discussion}

We have shown that under the assumptions of hot start core accretion, gas giants form with a radiative interior. For sufficiently large shock temperature, the accreting envelope is in the heating regime of \cite{Berardo2017}, and deposits material with entropy $S_{\rm min}$ at its base. For hot start boundary conditions, we find that $S_{\rm min}$ increases with time during accretion (see the orange trajectory in Fig.~\ref{fig: mins contour plot}). The entropy profile when accretion ends is set by the time-history of $S_{\rm min}$ (Fig.~\ref{fig: entropy_vs_mass}). Because $S_{\rm min}$ increases outwards, convection is inhibited and the interior is radiative.

The model we consider in this paper has $\eta=1$ (the hot start limit) so that all of the accretion energy is deposited in the planet (eq.~[\ref{eq:T0}]), but we find that radiative interiors from for a range of values of $\eta$. For $\dot M = 10^{-3}\ M_\earth\ {\rm yr^{1}}$ ($\dot M = 10^{-2}\ M_\earth\ {\rm yr^{-1}}$), models with $\eta\gtrsim 0.5$ ($\eta\gtrsim 0.05$) are in the hot regime. For lower values of $\eta$, the interior is fully convective as the envelope is able to match smoothly onto the interior adiabat during accretion (the stalling or cooling regimes of \citealt{Berardo2017}). We will explore the parameter space of $\dot M$ and $\eta$ in future work.

During cooling, convection penetrates from the surface into the interior. The planet eventually becomes fully-convective, at which point its internal entropy matches the initial entropy (here taken to be $\approx 9.5\ k_B/m_p$). The time to become fully convective is a few times shorter than the cooling time at this entropy (e.g.~Fig.~6 of \citealt{Marleau2014}), a timescale of $10^7\ {\rm yr}$ ($10^6\ {\rm yr}$) for a $10\ M_J$ ($1\ M_J$) planet (Fig.~\ref{fig: cooling profiles}).  At earlier times, the shape of the cooling curve is different from a traditional hot start, which is assumed fully-convective from the beginning (Fig.~\ref{fig: cooling curves}). The planet radius and luminosity are also smaller than a fully convective hot start with the same mass. Following the internal structure during formation is therefore crucial to make accurate inferences from direct imaging observations. For example, we find that at $\lesssim 10\ {\rm Myr}$ after formation, a $10\ M_J$ planet is less luminous than a traditional hot start by a up to a factor of 4. Since $L\propto M^2$ approximately (e.g.~\citealt{Marleau2014}), this translates to a derived mass larger by up to a factor of 2. The differences are smaller for lower masses and later times, e.g. tens of percent for $M=2\ M_J$ and ages $\gtrsim 10\ {\rm Myr}$.

Our results may have implications for the heavy element distribution in giant planets. Heavy elements are deposited in the envelope before runaway accretion begins \citep{Iaroslavitz2007,Helled2017}. Shutting down convection during accretion prevents mixing into the outer layers, confining heavy elements closer to the core. On the other hand, if planetesimals continue to deposit heavy elements as accretion proceeds (the extent to which this occurs is uncertain, e.g.~\citealt{HelledLunine2014}), they will not be mixed until after accretion when cooling begins. Depending on the distribution of heavy elements, the inwards growth of the convection zone may be suppressed, significantly delaying cooling (e.g.~\citealt{Leconte2012,Vazan2016}).

In this paper, we have focused on the thermal contribution to the stratification, and assumed a homogeneous composition. It will be interesting to incorporate the thermal stratification in models that compute the composition gradients in the evolution to cross-over mass and then runaway accretion phase (e.g.~\citealt{HelledLunine2014,Lozovsky2017}).  The continued accretion of planetesimals, and the resulting accretion luminosity deposited in the envelope could alter the thermal structure. We have also assumed here that the accretion rate and shock efficiency $\eta$ are constant during runaway accretion. Further investigations of the radiative transfer associated with the shock are needed to determine how the shock efficiency evolves. 

\acknowledgments
We thank A.~Burrows, G.-D.~Marleau and C.~Mordasini for helpful discussions. D.~B.~acknowledges support from a McGill Space Institute (MSI) Fellowship as well as a scholarship from the Fonds de Recherche Qu\'eb\'ecois sur la Nature et les Technologies (FQRNT). A.~C.~is supported by an NSERC Discovery grant and is a member of the Centre de Recherche en Astrophysique du Qu\'ebec (CRAQ). This work was partly carried out at the Aspen Center for Physics, which is supported by National Science Foundation grant PHY-1607611.

%\bibliographystyle{yahapj}
%\bibliography{references}

\end{document}